\documentclass[prl,twocolumn,showpacs,preprintnumbers,amsmath,amssymb]{revtex4}
\usepackage{dcolumn}
\usepackage{bm}
\usepackage{graphicx}
\usepackage{bm}
\begin{document}
\title{Observation of fluctuation induced tunneling conductance in polycrystalline CrO${_2}$}
\author{A. Bajpai}
\email{ashna@tifr.res.in}
\author{A. K. Nigam}
\affiliation{Tata Institute of Fundamental Research \\
Homi Bhabha Road, Colaba , Mumbai 40005 INDIA.}
\date{\today}
\begin{abstract}
    
   Intergranular conduction in half metallic  CrO${_2}$ is known to occur through a combination of spin dependent tunneling (driven by  Coulomb Blockade (CB) effects) together with certain spin independent (SI) hopping  processes. We present  evidence that in polycrystalline  CrO${_2}$ with  enhanced  grain size,  both these process ( CB effect and SI Hopping) are suppressed and the functional form of  conductance is best described by  Fluctuation Induced Tunneling (FIT) in a wide temperature range. Similar features are observed  when grain boundary density is increased by Cr${_2}$O${_3}$ or Cr${_2}$O${_5}$.The spin dependent tunneling driven by FIT results in the observation of significant  enhancement and monotonic temperature dependence of magnetoresistance. Overall, the magnetotransport measurements in a thus far unexplored crystallographic regime of CrO${_2}$ reveal that the functional form of conductance strongly influences its magnetoresistive properties. 
\end{abstract}

\maketitle 
CrO${_2}$ is a canonical example of a half metallic ferromagnet \cite{Groot} (HMFM), exhibiting an experimental observation of nearly 100 $\%$ spin polarization of the charge carriers at the Fermi level\cite{Soulen,Coey1, Dai1, Cheong1}. An interesting feature exhibited by this spintronic material is that electron transport is metallic in  single crystal form  whereas the granular samples  show an \emph{activated behavior}. It is to be noted that the bulk magnetization in both cases shows no apparent difference. This unusual behavior is understood to be the outcome of strong  grain boundary (GB) effects that governs the electronic conduction in this material\cite{Coey1,Dai1,Cheong1}. Here the inter grain conduction occurs through the (spin polarized) tunneling of  charge carriers  across  insulating  GB - a phenomenon which is  central to the observation of enhanced magnetorsistance (MR) in granular CrO${_2}$. \cite{Cheong1,Coey1,Dai1}  The GB's that appear as a thin layer on the surface of the CrO${_2}$ grain have been identified  to be an antiferromagnetic (AFM) insulator  Cr${_2}$O${_3}$.\cite{Coey1,Dai2}.  The typical thickness of the GB in commercially available CrO${_2}$ powders is about 1-2 nm whereas the typical grain size is about 0.1-0.4 $\mu$m \cite{Coey1,Dai2}.  The exact functional form of the conductance depends crucially on these microstructural parameters.

It is interesting to note that the transport mechanism as seen in  CrO${_2}$ as well as many such  granular metals simply mimic the physical properties of tailor made conducting  grains  (with well defined shape  and size ) embedded in an insulator\cite{abeles, sheng1,Coey1}. For instance, commercial powders are seen to exhibit Coulomb Blockade (CB) effects at temperatures below 50 K \cite{Coey1,Dai3}, with the conductance exhibiting a  characteristic exp(-$\Delta$/T)$^{1/2}$ dependence\cite{CB}. Above 50 K, multi channel inelastic hopping through the localized states in the GB  is seen to dominate the conductance. Overall, the functional form of conductance up to the room  temperature  can thus be written as \cite{Dai3}
  \begin{eqnarray}
\sigma = B_{1}exp(-\Delta/T)^{0.5} + C_{1}T^{1.33} + C_{2}T^{2.45} + ...
\end{eqnarray}   
   $C{_1}$, $C{_2}$ are the higher order terms is the coefficient corresponding to the first and second order  of hopping through the localized states of the GB. However, these hopping processes do not conserve spin and thus responsible for the rapid decline of spin polarization and consequently MR  above 50 K in this spintronic  material\cite{Dai3}. This functional form of conductance is seen to be retained when GB thickness is increased  by intentional dilution with Cr${_2}$O${_3}$. Though  enhancing GB thickness improves the MR at low temperatures, it still decays rapidly above 50 K and is negligible near the room temperatures\cite{Coey1,Dai3}.  Such studies  also indicate that the electron transport  and therefore MR  in this material should be tuned by variation in crystallographic microstructure. Unfortunately, tuning these parameters without disturbing the HMFM phase of CrO${_2}$ is a non trivial task, for it is  a metastable phase with associated difficulties in synthesis \cite{Chamberland}. These factors  severely limit the tunability factor of various  microstructural parameters and most magnetotransport measurements on this material have been reported on commercially available  powders, which were basically meant for magnetic recording industries. 
  
   In this letter, we report  electrical transport measurements in pure as well as highly diluted samples of  granular CrO${_2}$ with substantially enhanced grain size, roughly an order of magnitude larger than what is seen in  commercial powders. These samples were synthesized  in cold pressed  powders as well as sintered pellets form, while retaining the phase purity and characterized by X-ray diffraction (XRD) bulk magnetization and Scanning Electron Microscopy (SEM) etc. \cite{Bajpai1,Bajpai2}. The transport measurements were made using PPMS -9 (Quantum Design) by using the standard four probe technique in the temperature range of 5-320 K.  Fig. 1 summarizes the characterization measurements on a pure and sintered CrO${_2}$ sample, the  details of which  can be found elsewhere\cite{Bajpai2}. It is to be noted that we categorize  'pure CrO${_2}$' as  samples in which  saturation magnetization (Ms) falls between 125-135 emu/g\cite{note1}. These  samples do not exhibit any visible peak  corresponding to the insulating oxide Cr${_2}$O${_3}$ in XRD patterns. By suitable variation in synthesis condition, the mass fraction of Cr${_2}$O${_3}$ can be systematically increased, which is then visible in XRD and is  quantified\cite{Bajpai2}. We also present transport measurements on samples in which GB density was increased by Cr${_2}$O${_5}$, another AFM insulator having lower T${_N}$ than Cr${_2}$O${_3}$. The transport measurement on these composites bring out the crucial role played by the AFM character of GB in magnetotransport in granular CrO${_2}$, an aspect which is less adequately addressed in literature as far as bulk samples are concerned.  Exploring this material with varying crystallographic microstructure enables the observation of a different functional form of activated transport  which is evidently related with the significantly  enhanced MR properties in these samples.
  
Fig.2 displays zero field conductance as  a function of temperature for pure and diluted samples of CrO${_2}$.  The data set can be roughly divided into two temperature regions, the lower side of region I exhibits a practically linear temperature dependence, followed by a sudden rise in conductance occurs in region II (between 240-300 K), which was not seen in granular CrO${_2}$ prior to this work. Though conductance follows an activated behavior in both the regions, as is expected in granular CrO${_2}$, its temperature dependence  could not be fitted with either the CB term or its combination with  any other hopping terms as given in eq.1 \cite{Coey1,Dai3}in the temperature range of measurement.  It is to be noted that the CB energy has been calculated to be about 13 K \cite{Coey1} for commercially available CrO${_2}$ powders with  about  0.1  $\mu$m  grain size, whereas the grain size in our samples  is 1-10 $\mu$m \cite{Bajpai2}. Thus the absence of  CB effects in the lower temperature region is not surprising  due to the larger grain size \cite{CB}.  However, this data does not fit to any of the spin independent hopping described in eq.1 in  higher temperature region as well, in which the conductance rises much sharply and with a opposite curvature than what is predicted by higher order terms of Eq.1.
	
	  We find that the functional form of conductance in region I is best described on the basis of a Fluctuation Induced Tunneling (FIT) model meant for conducting grains separated by insulating barriers\cite{sheng2}. One essential ingredient of  this model is that the conducting grains  are large enough so as to overcome the Coulomb Blockade regime. The tunnel barrier is considered to be in the form of rectangular tunnel junction of area 'A' and width 'w'. When such a tunnel junction is kept under a thermal bath, apart from the externally applied field E${_A}$, there is an additional source of the field, which arises due to thermal fluctuations leading to a voltage difference across the junction. Most remarkably, it was seen that the underlying approximations, which are meant for a single tunnel junction can be used to describe a network of independently fluctuating tunnel junctions\cite{sheng2} and the final form of conductance  can be given by\cite{sheng2}.
\begin{eqnarray}
\sigma = \sigma_{0}exp(\frac{-T{_1}}{T+T{_0}})
\end{eqnarray}

here $T{_1} = uE_{0}^{2}/ k{_B}$, $T{_0} = 2uE{_0}^{2}/\pi \chi w k{_B}$, E${_0}$ = 4 V${_0}$/ew , u = wA/8$\pi$ , $\chi = (2m V_{0}/ h^{2})^{1/2}$ and V${_0}$ is the height of the potential barrier \cite{sheng2}. 

 The data set in Fig.2 fits well to Eq. 2 with single set of parameters  roughly upto 200 K . The ratio T${_1}$/T${_0}$ directly gives an estimate for barrier thickness if V${_0}$ is known. We estimate the parameter 'w' for the pure CrO${_2}$ by taking V${_0}$ to be 0.7 eV  \cite{Coey3} for CrO${_2}$. This  yields  'w' to be of  the order of 2A  which appears a feasible value for tunnel barrier thickness. Similar  fits were obtained on pure samples prepared in different batches in both cold pressed as well as in sintered form. Though FIT is  seen in granular  CrO${_2}$  samples for the first time, it is important to recall that is found to be most appropriate model for many  other granular oxide samples which exhibit intergranular tunneling. These include double pervoskite Sr${_2}$FeMoO6${_6}$ \cite{Fisher} and very recently Sr${_2}$CrWO${_6}$ \cite{Fisher2}.  
 
 Dilution \cite{Bajpai1, Bajpai2} of CrO${_2}$ with  insulating oxide Cr${_2}$O${_3}$  or Cr${_2}$O${_5}$  also follows a  similar functional form as is  observed for pure CrO${_2}$ in lower  temperature region.(Fig 2b).  Though functional form of conductance follows FIT even in highly diluted samples irrespective of the type of insulator- some subtle differences appear in region I. For instance, when the quantity (1/$\rho$)(d$\rho$/dT) is plotted as a function of temperature for pure  CrO${_2}$  and CrO${_2}$/Cr${_2}$O${_5}$ composites, all the data practically collapse into a single curve roughly upto 200 K  as is shown in the main panel of  Fig.3.  The CrO${_2}$/ Cr${_2}$O${_3}$ composites  do not follow this pattern as can be seen in the inset of Fig.3 where data on both composites is plotted together. This can be attributed to a strikingly different value of parameter T${_1}$  as seen in  Cr${_2}$O${_3}$ composites.  Considering the functional form of resitivity following Eq. 2, the quantity (1/$\rho$)(d$\rho$/dT) is given by  -T${_1}$/(T+T${_0}$)$^2$.  In the inset of Fig.3, the solid line is the function -T${_1}$/(T+T${_0}$)$^2$  for which T${_1}$ and T${_0}$ have been obtained from the fit of $\sigma$ vs T data for each composite. This separates out the region in which FIT dominate in a clear fashion and  brings out the robustness of the FIT conductance in pure and diluted CrO${_2}$ samples. Most importantly, it also gives a clue about the role of the magnetic nature of GB on the functional form of zero field conductance. Though  FIT model does not take care of magnetic interactions explicitly, its influence on the tunneling probability should reflect in the nature of the potential barrier, as it appears from the fitting parameters as obtained on samples with different type of GB.

	 The functional form of the conductance in  region II where a  sudden rise  occurs could not be explained by FIT model alone. As is evident from Fig2, the sharpness of 240 K feature varies with the type of GB. We observe that enhancing the GB density by  Cr${_2}$O${_5}$ reduces the sharpness of the step in conductance, however in the presence of Cr${_2}$O${_3}$ the functional form of conductance is qualitatively similar to what is seen in pure CrO${_2}$. It is to be noted that in the vicinity of the step, Cr${_2}$O${_3}$ is an AFM insulator whereas Cr${_2}$O${_5}$ is  a paramagnetic insulator. Though it  is difficult to separate out the contribution of grain or the GB from the present data, but it is  clear from Fig. 2 that the magnetic state of  GB does play a role.  The sudden rise in conductance  does not seem to be associated with a conventional structural or magnetic transition in CrO${_2}$ grain, as is concluded from the bulk magnetization and specific heat measurement around this temperature (Fig 1). However a subtle transition associated with magnetic GB can not be fully ruled out using these measurements alone and requires certain microscopic measurements. From the analysis of magnetoresistance data, it appears that a combination of  magnetic interactions among the grains and the GB is responsible for this feature.      
	 
 Fig 4 shows the effect of applied magnetic on the conductance for pure CrO${_2}$ in the entire temperature range of measurements. We find that  both  T${_1}$ and T${_0}$ change significantly in the presence of field though their ratio does not show variation.  We also find that both these parameters depend on the temperature and field  measurement protocols  but their ratio does not show significant variation on the cooling / heating cycles. Since the applied magnetic field aligns the magnetization of individual grains and also alters the magnetic state of GB, it is likely to  modulate the tunneling probability and the barrier height (V${_0}$) and hence influences T${_1}$ and T${_0}$. The value of V${_0}$ in the presence of the field  can be estimated by certain measurements  which is beyond the scope of present work \cite{Fisher}. We also note that the $\sigma$ Vs T curves remain well separated and exhibit a monotonic rise as a function of field till about 200 K. Above 240 K, the conductance rises sharply but the MR (defined as  ($\sigma$${_H}$-$\sigma$${_0}$)/ $\sigma$${_0}$) falls , though it  remains finite.  This is a significant observation for it shows that for samples with variation in crystallographic microstructure, finite MR can be retained right up to the room temperature. The data set in Fig 4 clearly brings out that the functional form of conductance plays a crucial role in determining the evolution of MR as a function of temperature.
 
	In conclusion we have demonstrated that by modulations in grain size, type and mass fraction of grain boundary density , we are able to observe a different functional form of activated transport in granular CrO${_2}$. In a wide temperature range , this is best described by Fluctuation Induced Tunneling irrespective of type and mass fraction of insulator.  These data convey that  tuning the microstructure, one can alter the functional form of activated transport which directly governs its magnetoresistive properties. The  magnetoresistance as measured in the sintered  samples of  CrO${_2}$ (and also when it is diluted with another insulating oxide) is substantially enhanced till about 200 K and is considerably large near the room temperature in these samples unlike vanishingly small MR as seen in commercial powders. These results should provide important input for fabricating the devices based on this spintronic material.
	
	 The authors thank  Prof. S.K.Malik for providing experimental facilities and Prof.  S. Bhattacharya for many insightful suggestions. AB gratefully acknowledges Dr.P. Chaddah for his support during her stay at CAT, Indore.

\newpage
\begin{figure}
\caption{Magnetization as a function of Temperature for pure and sintered  CrO${_2}$. Lower inset displayes SEM piture exhibiting grains with average length of about 5 microns.  Upper  inset  exhibits  specific heat as a function of temperature for the same sample.}
 
\caption{(a) displays $\sigma$ Vs T for pure CrO${_2}$.  The data is consistant with the previous observation of activated behavior for granular CrO${_2}$, but with a different functional form in the entire temperature range of measurement. The veritical line demarkate two distinct regions of activated trasport. Solid lines in main panel is the fit of FIT model (eq.2)  to the conductivity data. (b) displays the  same for CrO${_2}$/Cr${_2}$O${_5}$ (triangles) and CrO${_2}$/Cr${_2}$O${_3}$ composites (stars). The mass fraction of CrO${_2}$ in each of these composites is nearly 50 $\%$.}

\caption{The main panel shows the quantity (1/$\rho$)(d$\rho$/dT) as a function of temperature, for  pure  CrO${_2}$ and CrO${_2}$/Cr${_2}$O${_5}$ composites with varying mass fraction of CrO${_2}$. Inset shows the same plot for a CrO${_2}$/Cr${_2}$O${_3}$ (stars) together with a CrO${_2}$/Cr${_2}$O${_5}$ (triangles) composite. Solid lines in the inset  are the  function [-T${_1}$/(T+T${_0}$)$^2$] plotted against Temperature, where the corresponding  values of  T${_1}$ and T${_0}$  have ben  obtained  from the $\sigma$ vs T curves for each sample.}

\caption{ $\sigma$ Vs T measured at 0, 2 and 5 Tesla for a pure CrO${_2}$. Solid lines in each data set are the fit of eq.2.to the data.  The applied magnetic field does alters the parameters T${_1}$ and T${_0}$  though their ratio does not show varation. The deviation from FIT is also marked by the change in the temperature dependence of MR as is evident from the figure.}

\end{figure}

\begin{thebibliography}{ }
\bibitem[14]{note1} This value of saturation magnetization (Ms) is close to the theoretically predicted one (135 emu/g) and has been observed in granular samples  for the first time. The commercial powders exhibit Ms of the order of 110 emu/g.
\end{thebibliography}
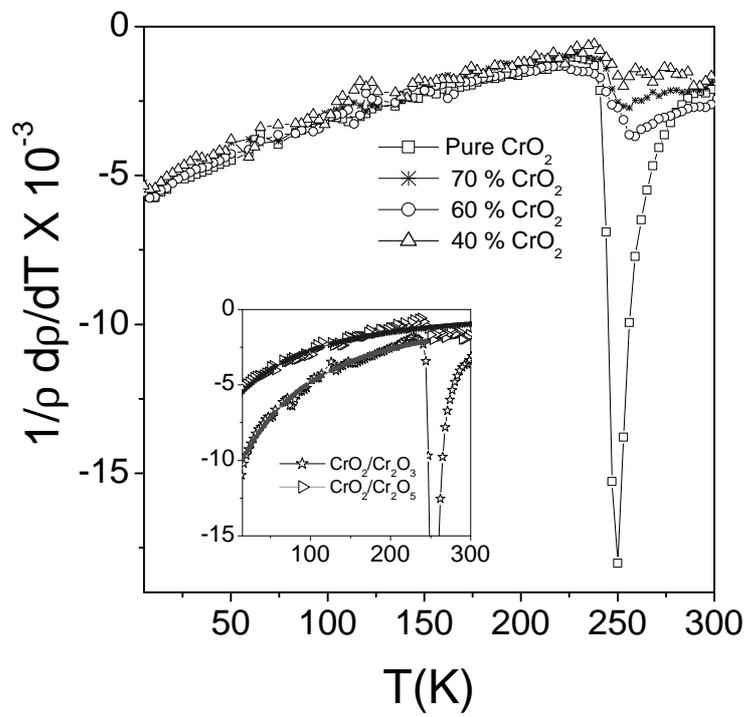
\end{document}